\documentclass[twocolumn,10pt]{article}

\usepackage{geometry}
\usepackage[utf8]{inputenc}
\usepackage{amssymb,amsmath}
\usepackage{graphicx}
\usepackage{caption}
\usepackage{layouts}
\usepackage{textgreek}
\usepackage[usenames,dvipsnames]{xcolor}

\usepackage{achemso}

\begin{document}

\twocolumn[
\begin{@twocolumnfalse}
\centering\Large{Voltage-time dilemma and stochastic threshold voltage variation in pure silver atomic switches}\vspace{0.3cm}

\noindent\large{Anna Ny\'ary,\textit{$^{a,b}$} Zolt\'an Balogh,\textit{$^{a,b}$} M\'at\'e Vigh,\textit{$^{a}$} Botond S\'anta,\textit{$^{a,b}$} L\'aszl\'o P\'osa,\textit{$^{a,c}$} and Andr\'as Halbritter$^{\ast}$\textit{$^{a,b}$}}\vspace{0.6cm}

\textit{$^{a}$~Department of Physics, Institute of Physics, Budapest University of Technology and Economics, Műegyetem rkp. 3., H-1111 Budapest, Hungary.}\\
\textit{$^{b}$~ELKH-BME Condensed Matter Research Group, Műegyetem rkp. 3., H-1111 Budapest, Hungary.}\\
\textit{$^{c}$~Institute of Technical
Physics and Materials Science, Centre for Energy Research, Konkoly-Thege M. \'ut 29-33., 1121 Budapest, Hungary.}\\
$^{\ast}$\textit{Corresponding author: halbritter.andras@ttk.bme.hu}\vspace{0.6cm}

\noindent\normalsize The formation and dissolution of silver nanowires plays a fundamental role in a broad range of resistive switching devices, fundamentally relying on the electrochemical metallization phenomenon. It was shown, however, that resistive switching may also appear in pure metallic nanowires lacking any silver-ion-hosting embedding environment, but this pure atomic switching mechanism fundamentally differs from the conventional electrochemical-metallization-based resistive switching. To facilitate the quantitative description of the former phenomenon, we investigate broad range of Ag atomic junctions with a special focus on the frequency-dependence and the fundamentally stochastic cycle-to-cycle variation of the switching threshold voltage. These devices are established in an ultra-high purity environment where electrochemical metallization can be excluded. The measured characteristics are successfully described by a vibrational pumping model, yielding consistent predictions for the weak frequency dependence and the large variance of the switching threshold voltage. We also demonstrate that electrochemical-metallization-based resistive switching and pure atomic switching may appear in the same device structure, and therefore the proper understanding of the pure atomic switching mechanism has a distinguished importance in silver-based electrochemical metallization cells.
\end{@twocolumnfalse} \vspace{0.6cm}]

\section*{Introduction}
In the recent decade, resistive switching phenomenon was established in various material systems, promoting novel, energy efficient, fast and compact technologies in the fields of nonvolatile data storage, in-memory computing or the hardware implementation of artificial neural networks.\cite{Yang2013,Ielmini2020,Wang2020,Lee2020,Li2021} A distinguished group of resistive switching memory devices relies on the voltage-controlled dissolution of metallic cations in an insulating matrix, resulting in the formation or destruction of ultra-small, close-to-atomic-sized metallic filaments between the contacting electrodes.\cite{Terabe2005,Valov2011,Valov2012,Yang2012,Geresdi2014,Gubicza2016,Choi2018,Wang2018,Santa2019, Santa2020, Wagenaar2012} In such structures the voltage-induced cation migration and the electrochemical metallization (ECM) play a key role in the programming of the devices, i.e. in the adjustment of the filament diameter.

Recently, it was shown, that resistive switching can also be induced in completely pure metallic wires, lacking any embedding, ion-hosting environment\cite{Martin2009,Schirm2013,Wang2016,Yoshida2017,Ring2020}. In this case reversible voltage-induced atomic rearrangements are observed in nanojunctions, where the size of the wire bottleneck approaches the ultimate single-atom regime. These atomic rearrangements are reflected by discrete jumps in the conductance state of the metallic nanowire. Such \emph{pure atomic switching} (PAS) was realized in various pure metallic systems, including Au, Cu, Al and Pb nanowires\cite{Martin2009,Schirm2013,Wang2016,Yoshida2017,Ring2020}. 

In this paper we argue, that the PAS phenomenon is not restricted to pure metallic nanowires, but it may become important in conventional electrochemical metallization cells as well, once the atomic-size-scales are reached. From this reason it has fundamental importance to explore the differences between PAS and ECM-type switching (ECMS), and to understand the physical mechanisms governing the former phenomenon. Though a broad range of material systems exhibit ECMS, silver plays a prominent role as an active material in such devices. To this end, we compare the PAS and ECMS characteristics in silver-based platforms, and to achieve an unquestionable distinction between the two types of phenomena, the PAS phenomenon is explored in ultra-pure silver nanowires, where electrochemical metallization can be excluded. Our analysis puts a special emphasis on the experimental investigation of the frequency dependence, as well the fundamentally stochastic variation of the switching threshold voltage, both being characteristic of the PAS phenomenon. Finally, we propose a vibrational pumping model of the PAS process, which consistently describes the observed phenomena.

\section*{Results and discussion}
\subsection*{Comparison of PAS and ECMS characteristics}
\begin{figure}[ht!]
	\includegraphics{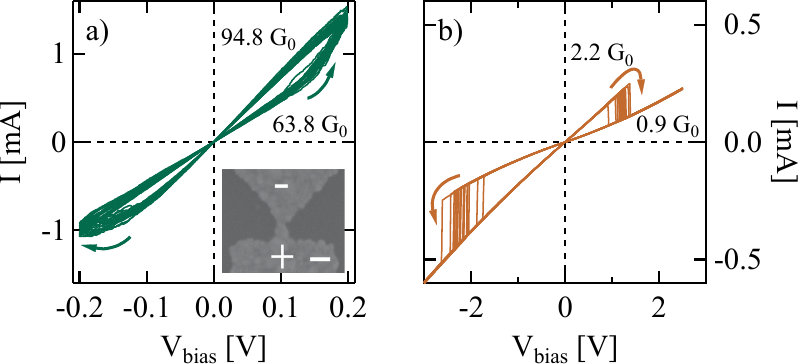}
	\caption{\it \textbf{Coexistance of ECMS and PAS characteristics in a nanofabricated Ag/Ag$_2$S/Ag memristive device.} Current-voltage characteristics of reproducible switchings measured in the nanoscale (a) and atomic-scale (b) regime, exhibiting ECMS (a) and PAS (b) characteristics. The legends demonstrate the average HCS and LCS conductances. These Ag/Ag$_2$S/Ag samples are fabricated by electron beam lithography establishing a $\approx 45\,$nm thick and $\approx 100\,$nm wide Ag nanobridge. This is further narrowed by a controlled electromigration protocol.\cite{Geresdi2010,Nef2014} The resulting nanojunction is exposed to vaporized sulfur to create the electrochemical metallization switching medium between the Ag electrodes. The inset of panel (a) shows a scanning electron microscopy picture of the constriction in the sample, the scale bar indicates 200\,nm distance. The $\pm$ signs indicate the voltage polarity on the junction at positive biasing. It is noted, that these devices lack the compositional asymmetry (Ag electrodes contact the junction from both sides), and therefore the well-defined switching polarity in case of ECMS is attributed to the geometrical asymmetry of the junction. In case of PAS the switching polarity varies according to the geometry of actual atomic arrangement of the junction.}
	\label{fig1}
\end{figure}

In our previous work, scanning tunneling microscopy point-contact measurements on sulphurized thin Ag samples already exhibited fundamentally different resistive switching characteristics in ultra-small atomic junctions compared to somewhat larger nanoscale point contacts.\cite{Geresdi2010} Here, we present another example of this phenomenon relying on our Ag/Ag$_2$S/Ag nanofabricated memristive devices\cite{Gubicza2016} (see the SEM image of the device in the inset of Fig.~\ref{fig1}a and more details in the caption). In the higher conductance regime, these nanojunctions were shown to exhibit conventional ECMS characteristics (see our previous results in Ref.~\citenum{Gubicza2016}), which is also illustrated by another dataset with 20 reproducible ECMS switching cycles in Fig.~\ref{fig1}a. In Fig.~\ref{fig1}b, however, we demonstrate a room temperature resistive switching of fundamentally different nature, highly resembling the PAS-type characteristics of other metallic species.\cite{Schirm2013,Wang2016,Yoshida2017} These switching characteristics were observed on the same Ag/Ag$_2$S/Ag nanofabricated memristive devices, by driving them to the conductance range of the G$_0=2e^2/h$ conductance quantum unit. Whereas the ECM-type switching of the larger junctions (Fig.~\ref{fig1}a) exhibit a gradual, and basically reproducible  transition between the high and low conductance states (LCS/HCS) at a few hundred mV switching threshold voltage characteristic of the Ag$_2$S resistive switching system,\cite{Gubicza2015,Gubicza2015a} Fig.~\ref{fig1}b exhibits a jump-like switching between two discrete atomic arrangements at an order of magnitude higher but fundamentally stochastic threshold voltage. In this illustrative measurement, however, the Ag junction was on purpose sulphurized, i.e. the pure metallic conditions were not satisfied, and the role of the silver-sulfide matrix in the atomic-sized switching process of Fig.~\ref{fig1}b remains unclarified. 

\begin{figure}[h!]
        \setlength{\belowcaptionskip}{-20pt}
        \includegraphics{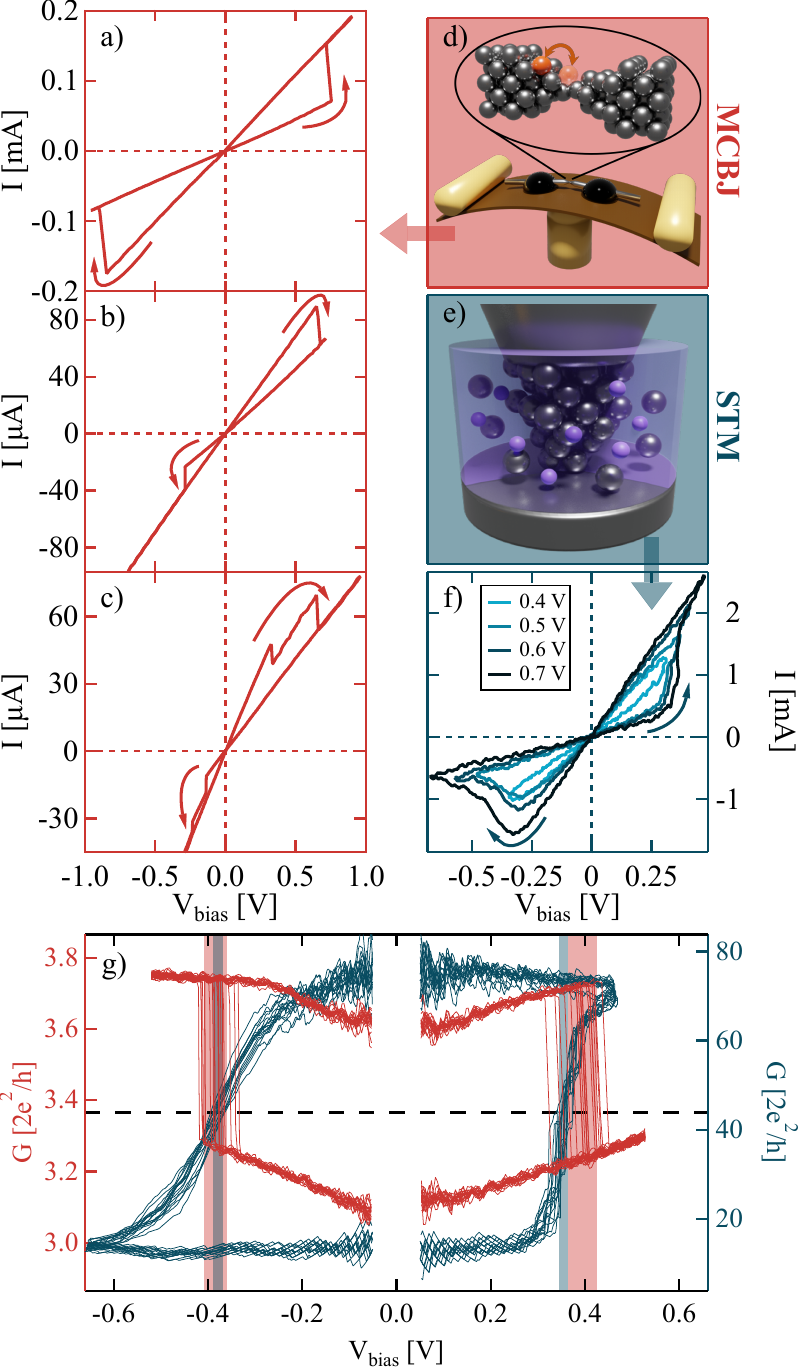}
	\caption{\it \textbf{Comparison of PAS and ECMS characteristics.} Panels (a,b,c) exemplify rather symmetric (a) or highly asymmetric (b,c) PAS characteristics with a reproducible switching between two (a,b) or three (c) atomic configurations. The PAS measurements are performed at cryogenic temperature ($T=4.2\,$K) using the MCBJ technique, which is illustrated in panel (d). In this case a serial resistance of 520 $\Omega$ was applied. As a comparison, ECMS is investigated on AgI thin films using room-temperature STM break junction technique (e), exhibiting driving-voltage-amplitude-dependent multi-level programming characteristics (f). The legends show the amplitudes of the driving triangular voltage signals. Here, a serial resistance of 150 $\Omega$ was applied. Panel (g) shows a representative PAS and ECMS with similar switching threshold voltages. Here, the conductance vs. bias voltage curve is plotted, and the black dashed line shows the mean value of the HCS and LCS conductances for for both systems. The switching threshold voltages and their standard deviations are evaluated at this conductance cut for both the set and reset transitions, yielding $\Delta V_\mathrm{thr}^\mathrm{set}/\overline{V}_\mathrm{thr}^\mathrm{set}=0.062$ and $\Delta V_\mathrm{thr}^\mathrm{reset}/\overline{V}_\mathrm{thr}^\mathrm{reset}=0.078$ for the PAS (red); while $\Delta V_\mathrm{thr}^\mathrm{set}/\overline{V}_\mathrm{thr}^\mathrm{set}=0.027$ and $\Delta V_\mathrm{thr}^\mathrm{reset}/\overline{V}_\mathrm{thr}^\mathrm{reset}=0.027$ for the ECMS (blue).}
	\label{fig2}
\end{figure}

Here, our primary goal is the investigation and the deeper understanding of the PAS mechanism in an environment, where any other types of resistive switching mechanisms can be ruled out. To this end, we have performed our measurements using a cryogenic temperature ($4.2\,$K) notched-wire mechanically controllable break junction (MCBJ) setup, where atomic-sized nanowires are established by the controlled mechanical rupture of a high purity metallic wire using a three point bending geometry (see the illustration in Fig.~\ref{fig2}d). The in-situ rupture in the cryogenic vacuum excludes the oxidation/contamination of the nanowire also ensuring an ultra-high mechanical stability. The current-voltage ($I(V)$) characteristics of the such-obtained ultra-high purity silver atomic wires readily exhibit typical PAS behavior (Fig.~\ref{fig2}a,b,c), resembling the PAS curves of other\cite{Schirm2013,Wang2016,Yoshida2017} previously studied but non-silver pure metallic systems.

Later on, we investigate further properties of the PAS mechanism in Ag atomic wires including the switching-to-switching and sample-to-sample variation of the $V_\mathrm{thr}$ switching threshold voltages as well as the sweep rate dependence of $V_\mathrm{thr}$. We compare these characteristics to the fundamentally different properties of conventional ECM-type resistive switching traces relying on the further analysis of our previously measured data on AgI resistive switching STM point contacts,\cite{Santa2020} where a significant amount of statistically independent data is available (see the illustration of the AgI point-contacts in Fig.~\ref{fig2}e and the example $I(V)$ curves in Fig.~\ref{fig2}f). As a next step, we propose a vibrational pumping model which is remarkably well describing the observed PAS characteristics. 

The different behavior of PAS and ECMS is already obvious from the switching $I(V)$ curves. (i) PAS displays abrupt jumps between discrete conductance states (Fig.~\ref{fig2}a,b,c) corresponding to discrete geometrical configurations of the atomic wire. As a sharp contrast, ECMS is accompanied by a more gradual transition between the high conductance state (HCS) and the low conductance state (LCS), as shown in Fig.~\ref{fig2}f. 
(ii) In case of ECMS the analog multilevel programming of the conductance states is a well-known phenomenon. This means, that the increase of the driving voltage amplitude continuously tunes the HCS and LCS conductances and thereby the $G_\mathrm{HCS}/G_\mathrm{LCS}$ conductance ratio (Fig.~\ref{fig2}f). \cite{Gubicza2015a,Santa2020}  In contrast, PAS yields driving-voltage-independent HCS and LCS conductances as long as a further atomic jump, resulting in discrete multilevel states is not reached (see Fig.~\ref{fig2}c for a PAS with three distinct conductance states). (iii) The switching direction (indicated by arrows in Fig. \ref{fig2}a,b,c,f) is defined by the material and geometrical asymmetry in case of ECMS, i.e. the set (LCS$\rightarrow$HCS) transition always happens at positive voltage polarity, where the the Ag layer is biased positively with respect to the electrochemically inert PtIr tip, and accordingly the reset (HCS$\rightarrow$LCS) transition occurs at negative bias. As a sharp contrast, in case of PAS the switching direction is determined by the local, atomic-sized geometry of the few-atom active region, and therefore the switching directions varies from junction to junction.

A further remarkable difference is observed in the cycle-to-cycle variation of the switching, as shown in Fig.~\ref{fig2}g demonstrating $\approx 20$ consecutive cycles for an example PAS (red) and ECMS (blue), both exhibiting set and reset transitions at similar voltages. 
Note, that in this case the $G=I/V$ conductance is plotted as a function of the bias voltage, and the switching threshold voltage is identified by the bias voltage, where the $G(V)$ curve crosses the average ($(G_\mathrm{HCS}+G_\mathrm{LCS})/2$) conductance (black bashed line). It is clear that the ECMS is a basically reproducible process, and accordingly the standard deviation of the 
switching threshold voltage, $\Delta V_\mathrm{thr}$ is small compared to the average value, $\overline{V}_\mathrm{thr}$. In contrast, the switching threshold voltage shows a broader, fundamentally stochastic cycle-to-cycle variation in case of PAS, yielding significantly higher $\Delta V_\mathrm{thr}/\overline{V}_\mathrm{thr}$ ratios (see the caption for the numerical values). 

\begin{figure}[ht!]
	\includegraphics{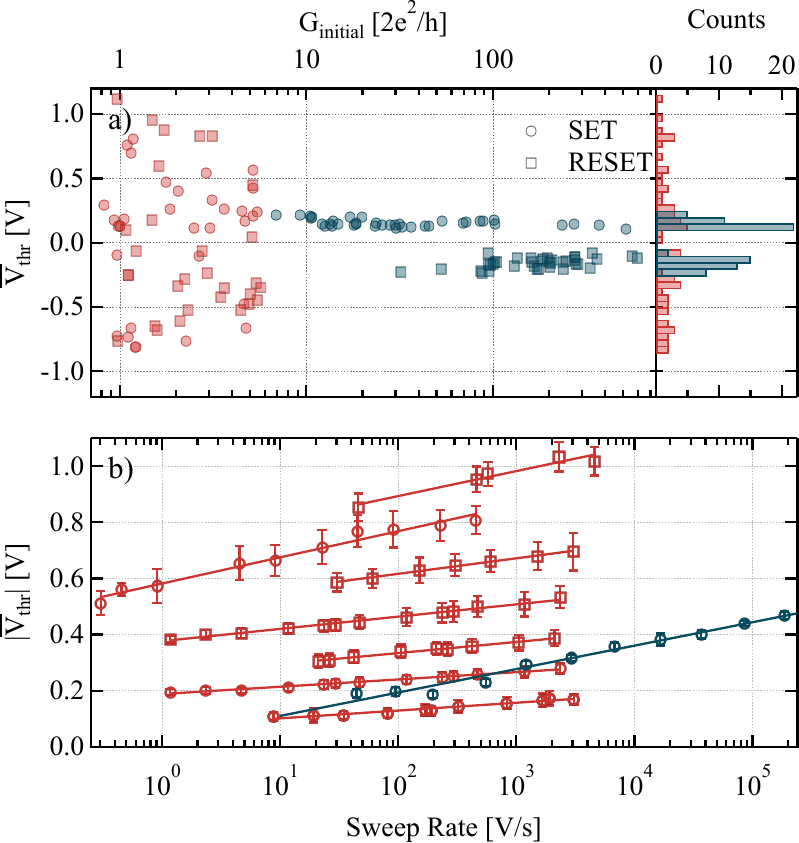}
	\caption{\it \textbf{Sweep-rate-dependence and device-to-device variation of the threshold voltage.} (a) The set (circle) and reset (square) threshold voltages of several PAS processes (red) and ECMS processes (blue) with various initial conductances. The left panel summarizes the data on a histogram. (b) Sweep-rate-dependence of the threshold voltages for representative PAS processes (red) and an ECMS measurement (blue). The error bars represent the cycle-to-cycle standard deviation of $V_\mathrm{thr}$ for the various measurements, the lines represent linear fits to the data along linear $V_\mathrm{thr}$ and logarithmic sweep rate axes.}
	\label{fig3}
\end{figure}

Next, we analyze the device-to-device variation of the threshold voltages. For this, in Fig.~\ref{fig3}a we plot the mean threshold voltages of SET transitions (circles) and RESET transitions (squares) for independent measurements of reproducible switchings as a function of the initial conductance state (i.e. the LCS/HCS in case of set/reset transition) both for PAS (red) and for ECMS (blue). The right panel of Fig.~\ref{fig3}a shows the corresponding histogram of the threshold voltages. In the figure, the difference between PAS and ECMS is clearly distinguishable: the threshold voltage in the former case shows a wide distribution exhibiting order-of-magnitude device-to-device differences, whereas the ECMS threshold voltages exhibit well-defined values. Moreover, we can observe again the clear confirmation about the switching direction being well-defined for ECMS (the set transition occurs at positive voltage) and stochastic for PAS. It is also noted, that the relation of the set and reset voltage also shows a broad variation in case of PAS, some curves exhibit similar set and reset voltages (see Fig.~\ref{fig2}a), while other junctions show extremely different set and reset voltages (see Fig.~\ref{fig2}b).

So far we have investigated the switching characteristics at specific frequencies, but now we turn to the time dependence, and aim to explore the so-called time-voltage dilemma, which is a well-known phenomenon in memristive systems,\cite{Waser2009,Gubicza2015a,Santa2020} and can be summarized as the exponential acceleration of the switching time while linearly increasing the applied voltage. Fig. \ref{fig3}b depicts the absolute value of the threshold voltage as the function of the sweep rate for selected representative atomic switchings in different ranges of threshold voltages (red symbols). A clear exponential dependence (i.e. linear trend on linear-logarithmic scale) is visible, however, the variation of the threshold voltage with the sweep rate is rather small: typically $V_\mathrm{thr}$ increases by $10\%$ as the sweep rate increases a decade (see the circles later in Fig.~\ref{fig4}d for a quantitative slope analysis). As a comparison, 
the Ag/AgI/PtIr ECMS system (blue curve) shows more than twice as large relative variation of $V_\mathrm{thr}$ in the same sweep rate interval, whereas there are also examples, where an Ag-based ECMS exhibits $\approx 100\%$ relative increase of the threshold voltage within one decade along the temporal axis.\cite{Tappertzhofen2012,VandenHurk2013,Chen2017}

\subsection*{Theoretical model of atomic switching and numerical simulations}
In the following we present a theoretical model with the goal of describing the PAS with special emphasis on the large stochastic cycle-to-cycle variation of the threshold voltage and the rather weak dependence of the threshold voltage on the sweep rate. Considering a conventional electromigration picture, the PAS could be related to current-induced forces, such that the current density is the relevant driving parameter of the switching\cite{Todorov2000}. In that case junctions with different diameters should perform atomic switching at similar current densities. This picture, however, was inconsistent with the experiments of Ref.~\citenum{Ring2020}, where atomic junction instabilities were studied for various materials and junction sizes. Instead, Ref.~\citenum{Ring2020} described the switching phenomenon by the pumping of particular phonon modes, also considering the contact destabilization by \emph{runaway phonon modes} due to non-conservative forces. Here, we apply a similar approach, but solely relying on a conventional vibrational pumping model,\cite{Todorov1998,Paulsson2005} i.e. excluding non-conservative forces. We demonstrate, that this simplified model is already sufficient to describe the sweep-rate dependence and the cycle-to-cycle variation of the threshold voltage.

\begin{figure}[ht!]
	\includegraphics{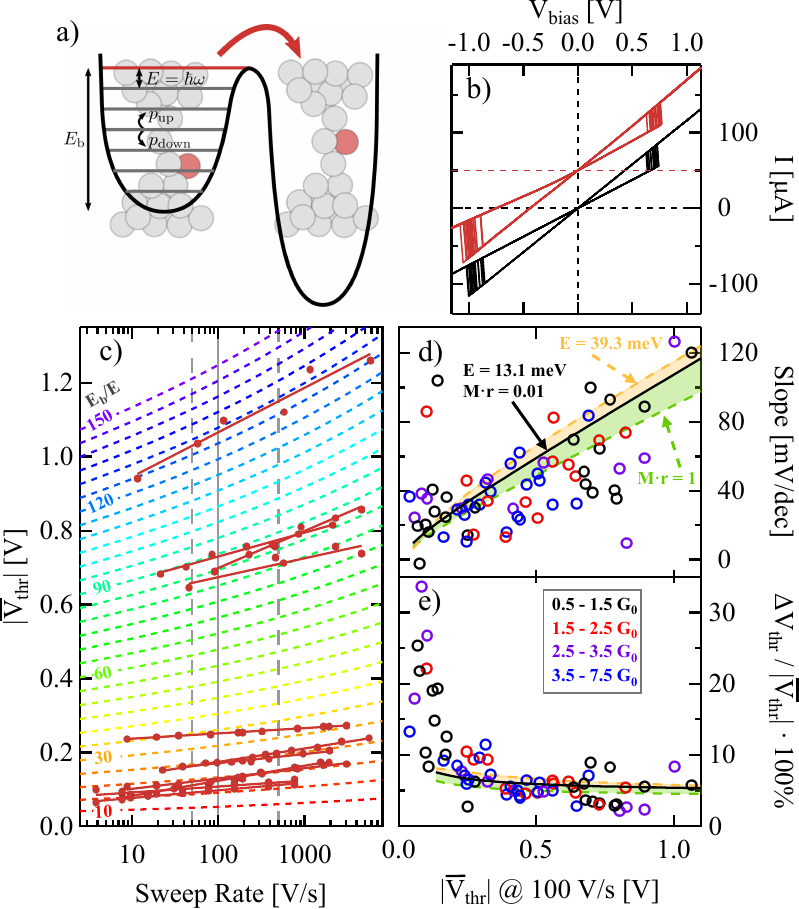}
	\caption{\it \textbf{Comparison of measured and simulated switching characteristics.} (a) Illustration of the model energy landscape and a cartoon naively illustrating the switching process. (b) Comparison of 20 measured (red) current-voltage characteristics and simulated (black) switching characteristics. The measured $I(V)$ curve is vertically shifted for clarity. (c) Measured (red dots) and simulated (colored dashed lines) sweep-rate-dependence of the threshold voltage. Both the simulations and the measurements are related to the case of a single conductance channel (i.e. junctions with initial conductance of $G=1\pm 0.15~G_0$ are selected in case of the measurements). (d) The slope of the simulated and measured $\overline{V}_\mathrm{thr}$ vs. $\log(\mathrm{Sweep Rate})$ curves as a function of $\overline{V}_\mathrm{thr}$ at $100\,$V/s sweep rate. The slope is evaluated in the $50-500\,$V/s interval (grey dashed lines in panel (c)). (e) The cycle-to-cycle standard deviation of the switching threshold voltage normalized to $\overline{V}_\mathrm{thr}$ as a function of the mean threshold voltage evaluated at $100\,$V/s. In panels (d) and (e) the colored circles represent the measured data according to the color-code in the legend of panel (e). The black lines represent the simulated data using $E=13.1$, $M\cdot r=0.01$ and $\gamma=3$. The yellow/green dashed lines respectively represent the variation of simulated results by changing a single parameter (i.e. using $E=39.3\,$meV/$M\cdot r=1$) and leaving all the other parameters unchanged.}
	\label{fig4}
\end{figure}

The basic idea of the model is presented in Fig.~\ref{fig4}a: two meta-stable junction configurations are considered, which are described by a double-well potential. At a certain voltage the final state is considered as the more stable (lower energy) configuration. However, an energy barrier of $E_\mathrm{b}$ must be reached for the switching. For this, the electrons cannot transfer enough energy to the switching atom within a single scattering process, however, multiple scattering events may \emph{pump} the proper vibrational mode, described by energy quanta $E=\hbar \omega$. Our model describes the voltage-induced evolution of the vibrational energy states as a random walk along the energy ladder, described by $p_\mathrm{up}\mathrm{(\Delta t)}$ and $p_\mathrm{down}\mathrm{(\Delta t)}$ upward and downward jump probabilities within a time-step $\Delta t$. These probabilities are taken from the zero temperature limit of the single vibrational level model of Ref.~\citenum{Paulsson2005} using $M$ open channels (see ESI for more details). This yields $$p_\mathrm{up}\mathrm{(\Delta t)}= \frac{2\cdot M \cdot r \cdot \Delta t}{h}\cdot\left( n+1 \right)\cdot \left( eV-E \right)$$
and 
$$p_\mathrm{down}\mathrm{(\Delta t)}= \frac{2\cdot M \cdot r \cdot \Delta t}{h}\cdot n \cdot \left( eV+\left( 3+4\gamma \right)\cdot E\right)$$ for the energy absorption from the electrons, or the energy emission from the vibrational mode to the electrons, respectively. Here $V$ is the bias voltage, $h$ is Planck's constant, $n$ is the occupation number of the vibrational mode and $r$ describes the ratio of the electrons interacting with the vibrational mode, which is approximated by the $r=0.01$ typical value according to the fitting of the point-contact spectroscopy measurements on an Au atomic contact in Ref.~\citenum{Paulsson2005}. 
The $\gamma=\gamma_\mathrm{d}/\gamma_\mathrm{eh}$ parameter describes the ratio of the electron-phonon coupling strength ($\gamma_\mathrm{eh}$) to the phonon-phonon damping parameter ($\gamma_\mathrm{d}$), as described in the ESI in more detail. 
In our simulations $\gamma=3$ is used based on the value determined for a gold wire in the work of Paulsson et al. \cite{Paulsson2005} The value of the phonon energy quantum is estimated to be $E=13.1$~meV in the Ag atomic wire based on the homology of forces\cite{Schober1983} applied to the simulated mean value of the mean phonon energy quantum of Cu atomic wires\cite{Ring2020}. According to the highly open conductance channels of Ag atomic wires, the number of open channels is estimated from the $G$ conductance as $M=\mathrm{round}(G/$G$_0)$. The model is invariant for the actual value of $\Delta t$, however the time-step should be sufficiently small to satisfy the $p_\mathrm{up}\mathrm{(\Delta t)}+p_\mathrm{down}\mathrm{(\Delta t)}<1$ condition for the investigated $n$ and $V$ values.

In our simulation the voltage is increased with a predefined sweep rate, and meanwhile the random walk along the energy ladder is considered according to the above rules (see the ESI for more details). The switching happens at the voltage, where the $E_\mathrm{b}$ barrier energy is reached for the first time. At this point the system switches to the other, more stable configuration, from which no back-switching is considered as the voltage is further increased. The simulation delivers the probability density function of the voltages needed to reach the $E_\mathrm{b}$ energy barrier at a given sweep rate. From this, both the average and the standard deviation of the switching threshold voltage can be determined at arbitrary sweep rate. 

We emphasize that our approach applies fixed, physically motivated model parameters, and the $E_\mathrm{b}$ barrier energy is the only free parameter, which is rather directly related to the $V_\mathrm{thr}$ switching voltage. On the other hand the energy landscape (Fig.~\ref{fig4}a) of the system is considered to be modified by the applied voltage, presumably yielding a decrease of $E_\mathrm{b}$ with increasing voltage\cite{}. Furthermore, the energetically more stable atomic configurations should be interchanged at reversed voltage polarity. These processes are not modeled in our simulation. We argue, however, that at low enough voltages, where the probability of reaching the barrier energy is negligible, the actual value of $E_\mathrm{b}$ is insignificant. Therefore, we consider $E_\mathrm{b}$ as the characteristic barrier height in the voltage interval, where the switching has a reasonable probability, i.e. in the voltage region, where the experimentally measured $V_\mathrm{thr}$ values scatter. We also emphasize, that the huge sample-to-sample variation of $V_\mathrm{thr}$ in Fig.~\ref{fig3}a implies a very broad variation of the possible atomic arrangements (and possible barrier energies). Accordingly, we cannot propose any specific atomic arrangements in our model, the cartoons in Fig.~\ref{fig4}a are only naive illustrations. 

\subsection*{Comparison of the experimental data and the model calculations.}

Fig.~\ref{fig4}b compares measured (red) and simulated (black) PAS $I(V)$ curves both including 20 switching cycles. In the simulation the HCS and LCS conductances and the sweep rate are set according to the measured $I(V)$ curve, and the $E_\mathrm{b}^\mathrm{set}$ and $E_\mathrm{b}^\mathrm{reset}$ barrier energies are adjusted such that the average switching voltages of the measurement are reproduced. The such simulated stochastic cycle-to-cycle variation of the switching voltage very well resembles that of the measured data.

To obtain a further comparison, we investigate the sweep-rate-dependence of the switching (Fig.~\ref{fig4}c). The simulated mean threshold voltages for $M=1$, $r=0.01$, $E=13.1\,$meV and $\gamma=3$ are plotted as a function of the sweep rate in Fig.~\ref{fig4}c for various $E_\mathrm{b}$ energy barriers by colored dashed lines. As a comparison the sweep-rate-dependent measured mean threshold voltages of PAS states with an initial conductance of $G=1\pm 0.15~G_0$ are shown by red circles. The red lines represent linear fits to the data along linear $V_\mathrm{thr}$ and logarithmic sweep rate axes. The slope of these fits very well resembles the slopes of the simulated curves, both exhibiting an increasing slope with increasing $V_\mathrm{thr}$. Note, that this comparison to the simulations allows the estimation of the energy barrier required for the atomic switching in a particular experiment (see the colored $E_b/E$ values on the simulated curves).

For a more quantitative analysis Fig.~\ref{fig4}d compares the threshold voltage dependencies of the measured and simulated slopes, i.e. the change of $\overline{V}_\mathrm{thr}$ along the one decade change of the sweep rate between the grey dashed lines in Fig.~\ref{fig4}c. 
Note, that in panel (b) the presented simulations and experiments were restricted to the single-channel ($M=1$) situation. Panel (d), however, shows the measured slopes (circles) for all the sweep-rate-dependent measurements as a function of $\overline{V}_\mathrm{thr}$ encoding the conductance range of the initial state with colors. This color-coding does not exhibit any obvious conductance dependence of the measured slope values. The $\overline{V}_\mathrm{thr}$-dependent simulated slopes at $M=1$, $r=0.01$ and $E=13.1\,$meV are plotted by black line. The colored dashed lines illustrate the sensitivity of the model to the parameters. A three-fold increase of $E$ (yellow dashed line) as well as a hundred-fold increase of $M$ or $r$ (green dashed line) yield a very modest variation of the simulated slope values. Note, that in the latter case the model only depends on the $M\cdot r$ product. Concluding this analysis: in spite of the scattering of the experimental data the simulated and the measured slope values cover the same slope range exhibiting a very similar $\overline{V}_\mathrm{thr}$-dependence. Furthermore, the simulated slope vs. $\overline{V}_\mathrm{thr}$ curve is rather insensitive to the model parameters, in particular the junction conductance ($G\sim M$) does not influence the slope values significantly.

Next, we investigate the $\overline{V}_\mathrm{thr}$ dependence of the cycle-to-cycle standard deviation of the threshold voltages, $\Delta V_\mathrm{thr}$ normalized to $\overline{V}_\mathrm{thr}$. The colored circles / solid and dashed lines are related to the same experimental / simulational data as in panel (d), using the same color-coding. This analysis also shows a very good correspondence between the experiment and the simulation, exhibiting a rather constant, $\Delta V_\mathrm{thr}/\overline{V_\mathrm{thr}}\approx 5\%$ value. A significant deviation from this is only observed at extremely low threshold voltages. In this range a few phonon energy quanta are sufficient to reach the barrier, i.e. the model calculation is expected to be less reliable. 

\section*{Conclusions}
In conclusion, we have performed a quantitative analysis of the pure atomic switching phenomenon in cryogenic temperature silver break junctions, where electrochemical-metallization-type resistive switching mechanisms can be excluded. We have demonstrated that PAS exhibits fundamentally different characteristics compared to conventional ECM-type resistive switching mechanisms, in particular, the switching threshold voltage exhibits a huge device-to-device variation, a large, fundamentally stochastic cycle-to-cycle variation, however, it depends rather weakly on the driving frequency. All these phenomena were successfully described by a simple model relying on the pumping of a single vibrational mode by the electrons scattering at the device bottleneck. Simulating the random walk along the energy ladder of the vibrational mode, and identifying the switching mechanism by the first occasion, when the proper barrier energy is reached, we could provide a consistent model of the experimental observations, quantitatively describing the weak frequency dependence of the switching threshold voltage, as well as its stochastic cycle-to-cycle variation. 

Our above measurements on undoubtedly pure Ag atomic wires very much resemble the switching characteristics of sulphurized silver devices driven to the range of the conductance quantum unit, i.e. Ag$_2$S resistive switching units with a few atoms in the active volume (see Fig.~\ref{fig1}b and Ref.~\citenum{Geresdi2010}). This comparison implies, that in the latter case the conventional ECMS is replaced by a fundamentally different physical mechanism, where the role of the Ag ions in the embedding Ag$_2$S environment become irrelevant, and rather the PAS phenomenon characteristic to atomic-sized pure metallic nanowires becomes the dominant process. This analysis raises the possible emergence of the PAS phenomenon in a wider range of atomic-sized ECM-type resistive switching systems, highlighting the need for the proper characterization and understanding of the PAS phenomenon. 

\section*{Acknowledgements}
This work was supported by the Ministry of Innovation and Technology and the National Research, Development and Innovation Office within the Quantum Information National Laboratory of Hungary, and the NKFI K128534, K143169 and K143282 grants. The support of the Hungarian Academy of Sciences in a form of Bolyai Janos Research Scholarship is also acknowledged by Z. Balogh and L. Pósa.

\section*{Author contributions}
The measurements on Ag nanowires and the evaluation of the experimental data were carried out by A. Nyáry. The modeling and the simulations were performed by Z. Balogh with fundamental contributions from A. Halbritter, A. Nyáry and M. Vigh. The measurements on Ag/AgI/PtIr samples were performed by B. Sánta, and the measurements on Ag/Ag$_2$S/Ag samples were performed by L. Pósa. A. Halbritter has supervised the work. The manuscript was written by A. Halbritter, A. Nyáry and Z. Balogh.

\bibliography{references}

\end{document}


\begin{center}\Large{Supporting Information:\\Voltage-time dilemma and stochastic threshold voltage variation in pure silver atomic switches}\vspace{0.3cm}

\noindent\large{Anna Ny\'ary,\textit{$^{a,b}$} Zolt\'an Balogh,\textit{$^{a,b}$} M\'at\'e Vigh,\textit{$^{a}$} Botond S\'anta,\textit{$^{a,b}$} L\'aszl\'o P\'osa,\textit{$^{a,c}$} and Andr\'as Halbritter$^{\ast}$\textit{$^{a,b}$}}\vspace{0.6cm}

\textit{$^{a}$~Department of Physics, Institute of Physics, Budapest University of Technology and Economics, Műegyetem rkp. 3., H-1111 Budapest, Hungary.}\\
\textit{$^{b}$~ELKH-BME Condensed Matter Research Group, Műegyetem rkp. 3., H-1111 Budapest, Hungary.}\\
\textit{$^{c}$~Institute of Technical
Physics and Materials Science, Centre for Energy Research, Konkoly-Thege M. \'ut 29-33., 1121 Budapest, Hungary.}\\
$^{\ast}$\textit{Corresponding author: halbritter.andras@ttk.bme.hu}\vspace{0.6cm}

\end{center}

\section*{Vibrational pumping model}

\begin{figure}[ht!]
	\center
        \includegraphics{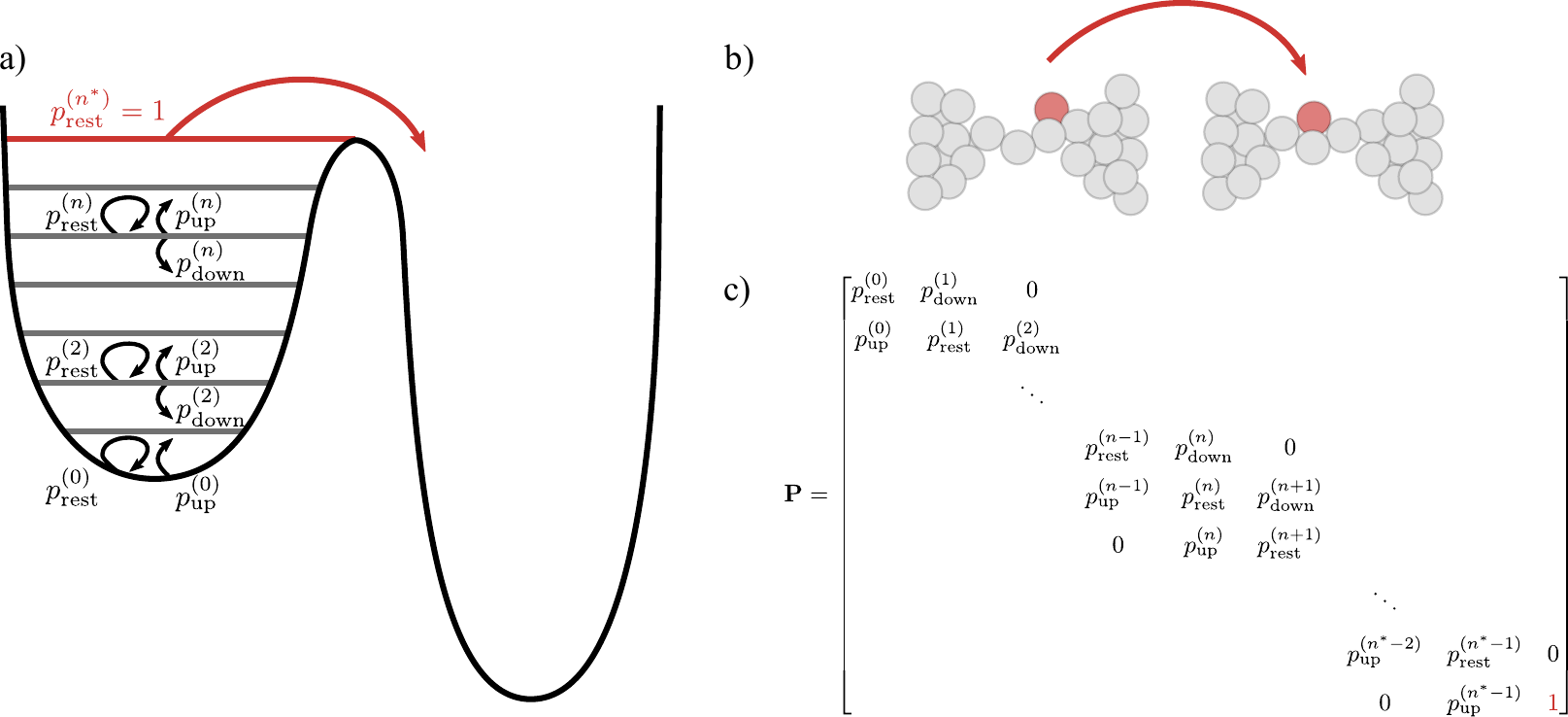}
	\caption{\it (a) Illustration of the model double-well potential with transition probabilities at different $n$ phonon occupation. As the phonon occupation reaches $n^*=E_\mathrm{b}/E$, the system switches into the lower energy configuration. (b) Naive picture of the atomic rearrangement occurring as the potential barrier is reached. (c) The stochastic matrix describing the transitions of a Markov chain. State $n^*$ is the absorbing state where $p_\mathrm{rest}^{(n^*)}=1$, leaving no transition possible out of this state.}
	\label{figESI1}
\end{figure}

In our simulation we rely on the single vibrational level model presented in the work of Paulsson \textit{et al.}\cite{Paulsson2005}, which gives the power dissipated by the electrons into a single vibrational mode as:
\begin{equation}
P=\gamma_\mathrm{eh}\cdot E\cdot \left[n_B(E)-n\right]+\frac{\gamma_\mathrm{eh}}{4}\cdot \frac{\left[\cosh{\left(\frac{eV}{kT}\right)}-1\right]\coth{\left(\frac{E}{2kT}\right)}\cdot E-eV\sinh{\left(\frac{eV}{kT}\right)}}{\cosh{\left(\frac{E}{kT}\right)-\cosh{\left(\frac{eV}{kT}\right)}}},
\label{eq:P}
\end{equation}
where $n_B$ is Bose-Einstein distribution, $n$ is the occupation number of the vibrational mode, $\gamma_\mathrm{eh}$ is the electron-hole damping rate (the electron-phonon coupling strength), and $E=\hbar\omega$ is the energy quantum of the vibrational mode. From this, the rate, at which the occupation number changes is obtained as: 
\begin{equation}
\nu=\frac{P}{E}+\gamma_\mathrm{d} \left[n_B(E)-n\right],
\label{eq:nu}
\end{equation}
where the $\gamma_\mathrm{d}$
external damping rate describes the  relaxation of the vibrational mode towards 
the phonon bath (phonon-phonon relaxation strength). In our calculation, we consider the zero temperature limit of this model, yielding 
\begin{equation}
\nu= \underbrace{\left[\frac{\gamma_\mathrm{eh}}{4\cdot E}\cdot \left(n+1\right)\cdot\left(eV-E\right) \right]}_{\nu_\mathrm{up}} - \underbrace{\left[ \frac{\gamma_\mathrm{eh}}{4\cdot E}\cdot n\cdot\left( eV+\left( 3+4\frac{\gamma_\mathrm{d}}{\gamma_\mathrm{eh}} \right)\cdot E\right)\right]}_{\nu_\mathrm{down}},
\label{eq:ndot}
\end{equation} 
where the formula is already decomposed for the excitation processes, where the occupation number increases ($\nu_\mathrm{up}$) or the relaxation processes, where it decreases ($\nu_\mathrm{down}$).
Our notation in the manuscript uses a dimensionless electron-phonon interaction rate, $r=\gamma_\mathrm{eh}\cdot h/8\cdot E$, and the $\gamma=\gamma_\mathrm{d}/\gamma_\mathrm{eh}$ normalized external damping rate, 
yielding the
\begin{equation}
    p_\mathrm{up}\mathrm{(\Delta t)} =    
    \frac{2 \cdot M \cdot r \cdot \Delta t}{h}\cdot\left( n+1 \right)\cdot \left( eV-E \right),
    \label{eq:pup}
\end{equation}
and
\begin{equation}
    p_\mathrm{down}\mathrm{(\Delta t)} = \frac{2 \cdot M \cdot r \cdot \Delta t}{h}\cdot n \cdot \left( eV+\left( 3+4\gamma \right)\cdot E\right),
    \label{eq:pdown}
\end{equation}
formulae for the upward and downward jump probabilities along the energy ladder (see Fig.~\ref{figESI1}). Note, that these probabilities already include a multiplication factor of $M$ generalizing the model for $M$ open conductance channels instead of the original single-channel situation. Lastly, we also consider the possibility of no event occurring while satisfying the relation 
\begin{equation}
    p_\mathrm{up}\mathrm{(\Delta t)}+p_\mathrm{down}\mathrm{(\Delta t)}+p_\mathrm{rest}\mathrm{(\Delta t)}=1.
    \label{eq:prest}
\end{equation}
Note, that the results of the simulation are insensitive to the $\Delta t$ timestep, however, $\Delta t$ should be sufficiently small to satisfy the  $p_\mathrm{up}\mathrm{(\Delta t)}+p_\mathrm{down}\mathrm{(\Delta t)}<1$ condition. In our simulation, $\Delta t$ was typically in the range of $10^{-14}-10^{-17}\,$s.

In Ref.~\citenum{Paulsson2005} the single vibrational level model was successfully applied to describe the experimentally measured nonlinearities in the current voltage ($I(V)$) characteristics of single-atom and single-molecule junctions due to the scattering of the electrons on a vibrational mode. For a single-channel perfectly transmitting junction the differential conductance exhibits a jump-like decrease from the $dI/dV=1\,$G$_0$ low bias value as the voltage reaches the vibrational quantum ($eV=E$), and along the further elevation of the voltage, the differential conductance further decreases according to the increasing electron-phonon scattering probability, and the increase of the $n$ average occupation number. In our model the applied $\gamma=3$ and $r\sim 0.01$ parameters are obtained from the
$\gamma_\mathrm{d}$ and $\gamma_\mathrm{eh}$ values of the single vibrational level model best fitting the voltage dependent differential conductance curves of single-atom-wide Au nanojunctions.\cite{Paulsson2005} It is noted, that in the oversimplified limit, where a single-channel perfectly transmitting atomic junction is considered at $T=0$ and $\gamma_\mathrm{d}=\infty$ (i.e. the vibrational mode is always in the ground state), the $r$ parameter of our model directly describes the magnitude of the differential conductance decrease at $eV=E$, i.e. in this oversimplified situation the applied $r=0.01$ parameter would express a $1\%$ decrease of the differential conductance at $eV=E$. In a perfectly transmitting channel the electrons can only scatter backwards, i.e. a $1\%$ decrease of the differential conductance corresponds to the scattering of $\approx 1\%$ of the electrons on the vibrational mode.

\section*{Simulation}

In our model we consider the random walk along the phonon occupation ladder (pictured in Fig.~\ref{figESI1}a) and assign the bias-dependent probabilities expressed in Eqs. \eqref{eq:pup}-\eqref{eq:prest} to the transitions. We identify the atomic switching event with the first occasion, where the $n\cdot E$ vibrational energy reaches the $E_\mathrm{b}$ barrier height, at that moment a sudden atomic rearrangement is considered, as illustrated in Fig.~\ref{figESI1}b. 
Note that the random walk of the consecutive transition events with state-dependent probabilities is actually an example of a Markov chain model.\cite{KemenySnell1976} In this particular case, the states are indexed by the $n$ occupation number, and the Markov chain model can be described by the the bias-dependent stochastic matrix, outlined in Fig.~\ref{figESI1}c, where $P_{j,i}$ describes the transition probability from state $n=i$ to $n=j$.  Accordingly the $P_{n+1,n}$, $P_{n-1,n}$ and $P_{n,n}$ components are respectively identified by the $p_\mathrm{up}^{(n)}$, $p_\mathrm{down}^{(n)}$ and $p_\mathrm{rest}^{(n)}$ probabilities. The final state ($n=n^*$) where $n^*\cdot E=E_\mathrm{b}$ is satisfied, is an absorbing state, ($P_{n^*,n^*} = p_\mathrm{rest}^{(n^*)} = 1$) from which the system cannot jump to any other state. We describe the system with the $v_i(k)$ state vectors describing the probability that at time $t=k\cdot \Delta t$ the system stays in the state $n=i$. At $t=0$ ($k=0$) the system is started from the ground state ($v_{i=0}(k=0)=1$, $v_{i\ne 0}(k=0)=0$). The evolution of the state vector is described by the $v_j(k)=\sum_i{P_{j,i} \cdot v_i(k-1)}$ equation. Finally, we are interested in the  absorption probability (i.e. the probability that the state $n^*$ is reached at time-step $k$, or in other words the system exhibits an atomic switching at time-step $k$) which is given by $p_\mathrm{switch}(k)=v_{n^*}(k)-v_{n^*}(k-1)$.

\begin{figure}[ht!]
	\center
        \includegraphics{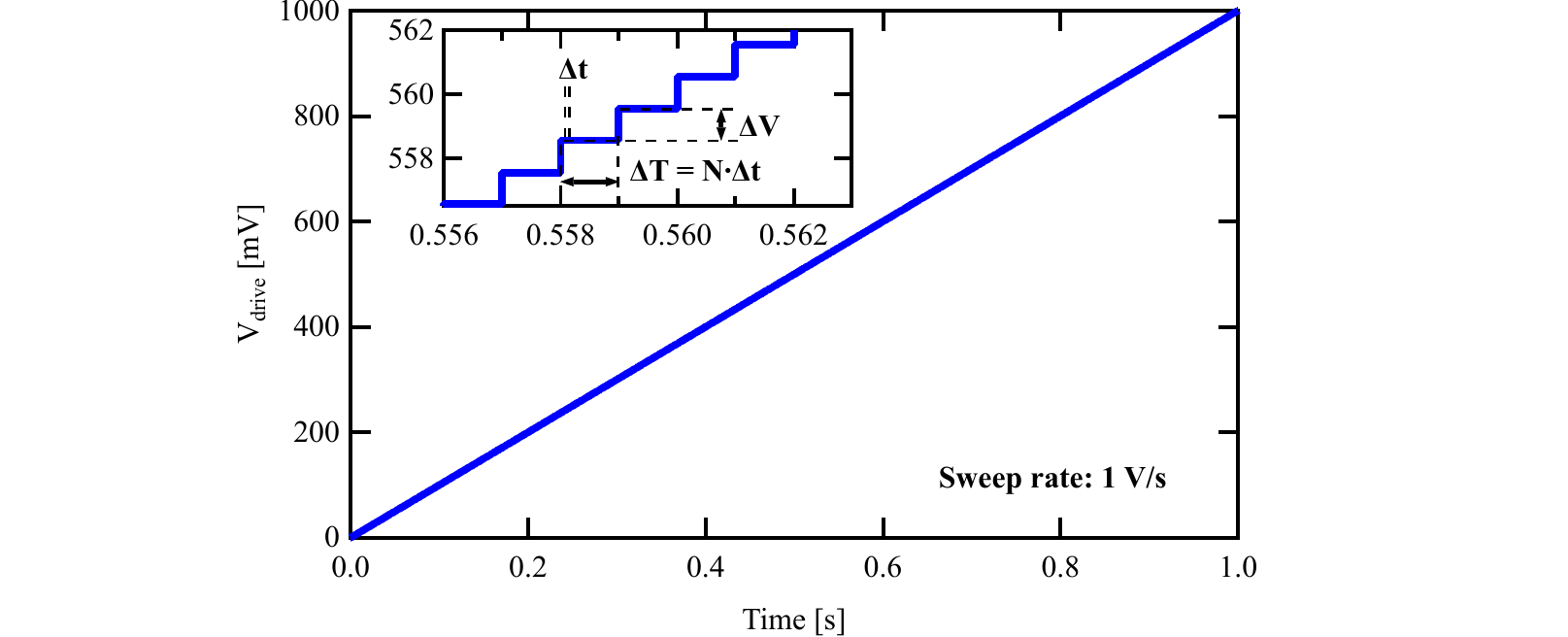}
	\caption{\it Simulated triangular voltage signal at the sweep rate of $\Delta V/\Delta T=1$~V/s. The voltage signal is discretized with steps of $\Delta V = 0.1\,$mV yielding $\Delta T=N\cdot\Delta t$ long plateaus where $\Delta t$ is the defined elementary time-step.}
	\label{figESI2}
\end{figure}

\begin{figure}[ht!]
	\center
        \includegraphics{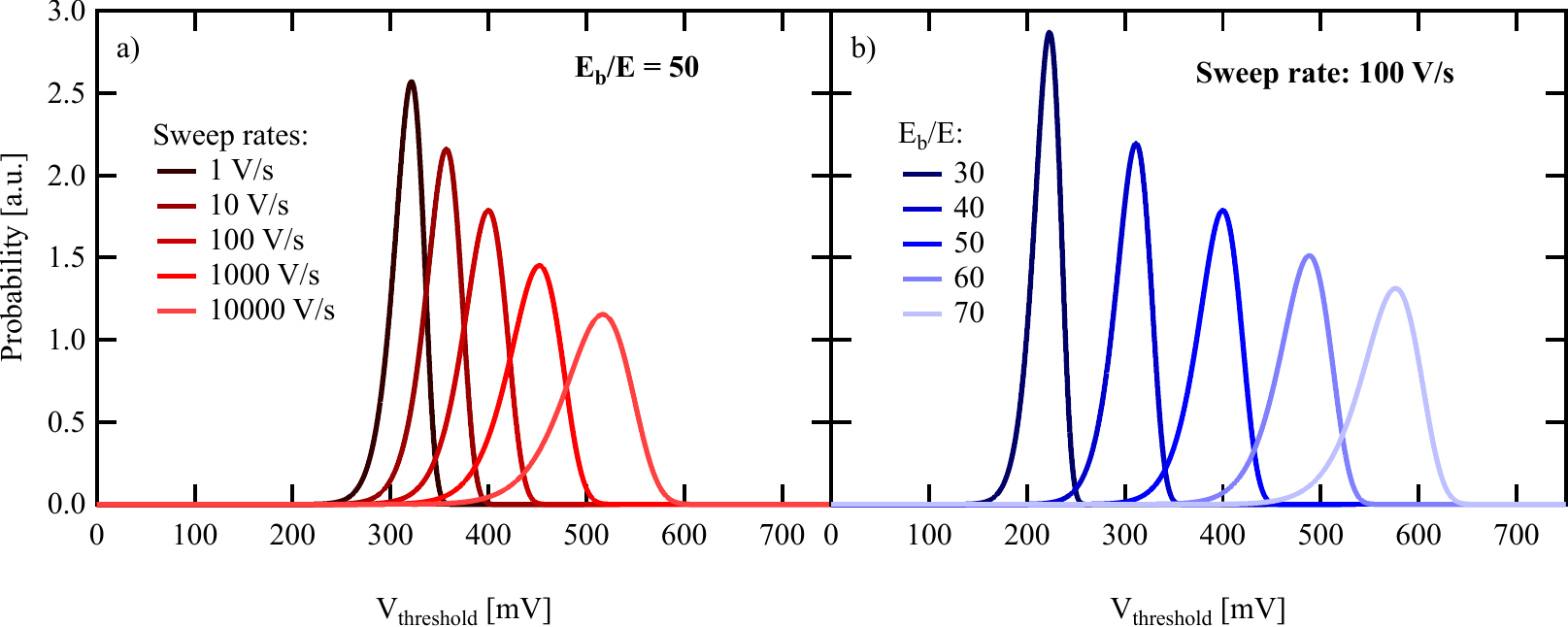}
	\caption{\it Probability density functions of the threshold voltages at fixed barrier height $E_\mathrm{b}/E=50$ for different sweep rates (a) and at fixed sweep rate of 100~V/s for different barrier heights (b).}
	\label{figESI3}
\end{figure}

To simulate the PAS $I(V)$ curves, we consider a step-wise increase of the $V_\mathrm{drive}$ driving voltage using $\Delta V = 0.1\,$mV voltage steps, where the actual voltage is obtained as $V=M\cdot \Delta V$, $M$ being the integer step number. Consequently, the sweep rate is obtained as $\Delta V/\Delta T$, where $\Delta T=N\cdot \Delta t$ is the length of each voltage step (see Fig. \ref{figESI2} for the simulated voltage sweep). To speed up the simulation, we do not capture the actual switching time with $\Delta t$ resolution, we just identify the voltage step at which the switching happens. Note, that at the end of voltage step $M$ the state vector $v_{j}^{(M)}$  can be obtained from the state vector at the end of the previous voltage step as $v_j^{(M)}=\sum_i{\left(P^{N}\right)_{j,i} \cdot v_i^{(M-1)}}$, where the $N={\Delta T/\Delta t}$ power of the $P_{j,i}$ probability matrix describes the number of elementary $\Delta t$ time-steps along one voltage step. Utilizing this property, we evaluate the proper powers of the probability matrix at each discrete voltage value. Finally, we determine the probability density function $p_\mathrm{switch}(V)=v_{n^*}^{(M)}-v_{n^*}^{(M-1)}$ describing the probability that the switching will happen at the $V=M\cdot \Delta V$ voltage-step. From this, both the mean switching voltage and the standard deviation of the switching voltage can be determined.  Fig. \ref{figESI3} demonstrates the resulting probability density functions. As a general observation, both an increased sweep rate and an increased barrier height broaden the distribution of the threshold voltages while also increasing the mean value.

\bibliography{references_SI}